\documentclass[%
aip, 
apl,
reprint,
superscriptaddress,
 amsmath,amssymb,
floatfix,
]{revtex4-1}

\usepackage{graphicx}
\usepackage{dcolumn}
\usepackage{bm}
\usepackage{color}

\begin{document}

\preprint{APS/123-QED}

\title{Towards wafer-scale diamond nano- and quantum technologies}

\author{Richard Nelz}
\author{Johannes G\"orlitz}
\author{Dennis Herrmann}
\author{Abdallah Slablab}
\author{Michel Challier}
\author{Mariusz Radtke}
\affiliation{Universit\"at des Saarlandes, Fakult\"at NT, Physik, 66123 Saarbr\"ucken, Germany}%
\author{Martin Fischer}
\author{Stefan Gsell}
\affiliation{Augsburg Diamond Technology GmbH, 86159 Augsburg, Germany}
\author{Matthias Schreck}
\affiliation{Universit\"at Augsburg, Institut f\"ur Physik, 86135 Augsburg, Germany}
\author{Christoph Becher}
\author{Elke Neu}
\email{elkeneu@physik.uni-saarland.de}
\affiliation{Universit\"at des Saarlandes, Fakult\"at NT, Physik, 66123 Saarbr\"ucken, Germany}%

\date{\today}


\begin{abstract}
We investigate native nitrogen (NV) and silicon vacancy (SiV) color centers in commercially available, heteroepitaxial, wafer-sized, mm thick, single-crystal diamond. We observe single, native NV centers with a density of roughly 1 NV per $\mu$m$^3$ and moderate coherence time ($T_2 = 5\,\mu$s) embedded in an ensemble of SiV centers. Low-temperature spectroscopy of the SiV zero phonon line fine structure witnesses high crystalline quality of the diamond especially close to the growth surface, consistent with a reduced dislocation density. Using ion implantation and plasma etching, we verify the possibility to fabricate nanostructures with shallow color centers rendering our diamond material promising for fabrication of nanoscale sensing devices. As this diamond is available in wafer-sizes up to 100\,mm it offers the opportunity to up-scale diamond-based device fabrication. 
\end{abstract}

\pacs{Valid PACS appear here}
\maketitle
\begin{figure}
	\centering
	\includegraphics[width=1\linewidth]{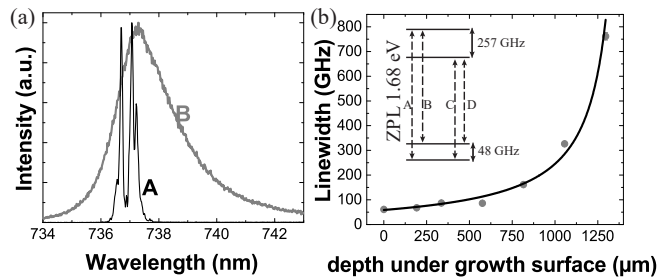}
	\caption{(a) PL spectra of the SiV ensemble at $10\,$K. Close to the growth surface, a clear fine structure is visible (black curve \textit{A}) while the ZPL continuously broadens along the growth direction. Closer to the substrate, broadening completely masks the fine structure components (grey curve \textit{B}). (b) Linewidth of the transition C along the growth direction of the diamond. Inset: SiV level scheme according to Ref.\ \onlinecite{Arend2016}}
	\label{fig:SiV}
\end{figure}
\begin{figure}
	\centering
	\includegraphics[width=1\linewidth]{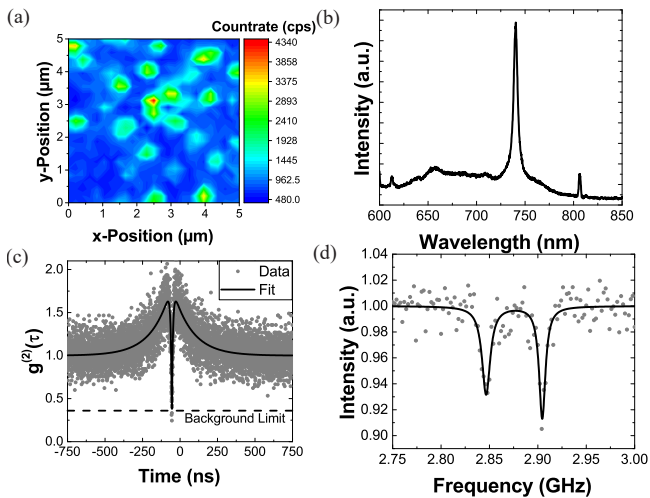}
	\caption{(a) PL map of the diamond (detected wavelength range $680\,$nm to $720\,$nm) showing $\approx$ 25 bright spots which all have NV PL. (b) Room temperature PL spectrum of one of these NVs. Bright SiV ZPL is visible leading to the background measured in the PL map in (a) despite spectral filtering. The narrow PL line observed at 805 nm is most probably connected to the SiV ensemble.\cite{Neu2011,Lindner2018} (c) $g^{(2)}$-measurement of one of the NVs. A clear anti-bunching is visible. $g^{(2)}(0)$ is only limited by the background of the SiV ensemble (see supplementary material). (d) ODMR measurement of single NV showing a background-corrected contrast C above $10\,\%$ (external magnetic field approx.\ $11\,$G).}
	\label{fig:NVsingle}
\end{figure}
\begin{figure}
	\centering
	\includegraphics[width=1\linewidth]{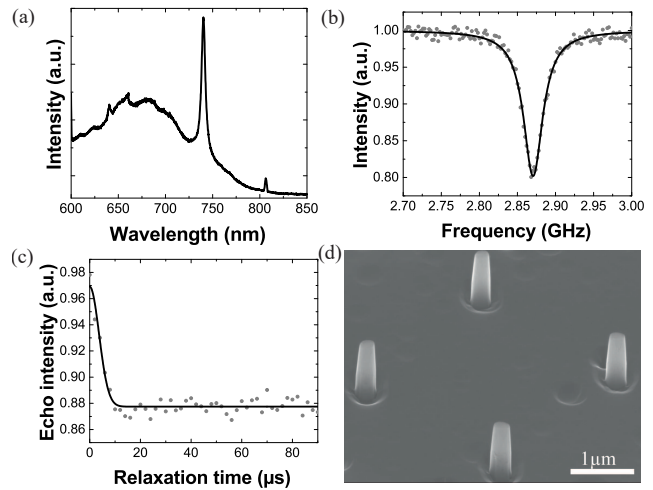}
	\caption{(a) PL spectrum of the implanted NV ensemble. A clear NV$^-$ as well as minor NV$^0$ spectrum is visible while SiV PL remains unchanged. (b) ODMR spectrum without external magnetic field strongly broadened due to strong microwave pumping of the transition (unbroadened linewidth $10.67\,$MHz, $C=20\,\%$). (c) Spin-Echo-measurement of the implanted NV centers, fitted employing the formalism derived in Ref. \onlinecite{Childress2006}. The coherence time is $T_2 = 5.2(3)\,\mu$s. (d) Scanning electron microscopy image of nanopillars etched into the growth surface of the diamond. The cylindrical pillars have a diameter of $\approx 200\,$nm and a height of $\approx 1\,\mu$m.}
	\label{fig:NVensemble}
\end{figure}

Diamond nanostructures are of significant importance for various applications in science and industry including nanomechanical devices,\cite{Teissier2014} photonics \cite{Aharonovich2014a} and sensing.\cite{Rondin2014} A major challenge for most of these applications is the scalability of the fabrication process predominantly due to a lack of large area single-crystal diamonds with good crystalline quality and high purity. Manufacturing synthetic, single-crystal diamond on wafer-scale has been an active field of research \cite{Yamada2014,Schreck2017} leading to the commercial availability of single-crystal diamonds with a diameter of $\approx 100\,$mm recently. This progress opens the road towards up-scaling the fabrication of single-crystal nanostructures especially for diamond related sensing applications.\cite{Appel2016} Color centers in diamond, in particular the negatively-charged nitrogen vacancy (NV) center in nanostructures, have been extensively used to sensitively measure e.g.\ magnetic fields in the last decade.  Recently, silicon vacancy (SiV) centers emerged as alternative enabling all optical sensing of temperatures using their narrow electronic transitions.\cite{Ngyuen2018}  Single color centers allow for sensing with high spatial resolution and offer bright, photostable photoluminescence (PL). In addition, NV centers provide highly-coherent, controllable spin states \cite{Rondin2014} and show optically-detected magnetic resonance (ODMR) enabling to read out their spin states via PL detection. As a consequence, even single NV centers can serve as quantum-enhanced sensors. Magnetic field imaging using NV centers has various applications ranging from material characterization in superconductors\cite{Thiel2016} or magnetic materials for spintronics\cite{Tetienne2014} to life science applications where nuclear magnetic resonance detection of single proteins is of interest.\cite{Lovchinsky2016}\\
We here demonstrate the basic applicability of commercial, single-crystal, wafer-sized diamonds for quantum technology applications. To this end, we demonstrate coherent manipulation of single native NV center spins in the material, while low-temperature spectroscopy of SiV center PL indicates high crystalline quality of the material. In addition, we implant shallow NV centers (depth $\approx 10\,$nm), indispensable for nanoscale sensing, and investigate their suitability for quantum sensing applications. Furthermore, we apply reactive ion etching to the material to illustrate routes towards up-scaling of diamond-related nanofabrication. \\
We use (100) oriented single-crystal diamonds from AuDiaTec (Augsburg Diamond Technology GmbH) synthesized in a microwave plasma enhanced chemical vapor deposition process based on a methane/hydrogen (CH$_4$/H$_2$) gas mixture at a temperature $>1000\,^\circ$C. H$_2$ was purified by passing it through a AgPd membrane, while the used CH$_4$ had a purity of 99.9995\%. As a consequence, we estimate the nitrogen (N$_2$) concentration in the gas phase to be $<0.5\,$ppm.  The Ir/YSZ(yttria stabilized zirconia)/Si substrate was removed by chemical etching and grinding before the diamond wafer was cut into pieces of $10\,$mm$\times$$10\,$mm using a pulsed IR laser. The diamonds are $1.3\,$mm thick. For our experiments, we use the crystals with as grown surface. Diamond synthesis at AuDiaTec is based on the technology developed at the University of Augsburg. As described in Ref.\ \onlinecite{Schreck2017}, successful scaling of heteroepitaxial nucleation and growth to $100\,$mm wafer-size was achieved by the use of the multilayer substrate Ir/YSZ/Si. During heteroepitaxial growth of thick (mm) single crystals, the dislocation density decreases proportional to the inverse of the diamond layer thickness indicating high crystalline quality for this material at the growth surface.\cite{Schreck2017}\\
To investigate PL and perform spin manipulation of NV centers, we use a home-built confocal microscope (Numerical aperture NA 0.8, $532\,$nm laser excitation) where confocal filtering of the PL is ensured by using a single mode optical fiber. To quantify the PL intensity, we use highly-efficient photon counters (Excelitas SPCM-AQRH-14, quantum efficiency $\approx 70\,\%$). In addition, we use a grating spectrometer (Acton Spectra Pro 2500, Pixis 256OE CCD) to record the PL spectrum. To measure PL lifetimes, we employ time correlated photon counting (PicoQuant, PicoHarp 300) and pulsed laser excitation (NKT EXW-12, pulse length $\approx 50\,$ps, wavelength $527$-$537\,$nm). The setup is equipped with a microwave source and amplifier (Stanford Research Systems, SG384 and Mini Circuits, ZHL-42W+) for spin-manipulation of NV centers. To investigate SiV centers, we use a second confocal microscope (Numerical aperture NA 0.9) embedded into a liquid helium flow cryostat employing a titanium-sapphire laser (Matisse, TX) at $690\,$nm as excitation source. For confocal filtering, a single mode optical fiber is used. PL spectra of SiV centers are recorded using a grating spectrometer (Jobin Yvon, Horiba, Grating: $1800\,$g/mm).\\
We now discuss the PL results from the diamond under continuous laser excitation at $532$\,nm. Close ($< 5\,\mu$m) to the nucleation side, we observe very bright PL consisting of a broad background as well as distinctive NV and SiV PL.  In contrast, PL maps recorded in a wavelength range from $680\,$nm to $720\,$nm from planes parallel and closer to the growth surface show  bright, distinguishable spots [see Fig. \ref{fig:NVsingle}(a)]. The spots clearly show PL spectra of NV centers [see Fig. \ref{fig:NVsingle}(b)], constituting the first observation of individual native NV centers in heteroepitaxial single crystal diamond. Additionally, these PL spectra reveal the presence of homogeneously distributed SiV centers in the sample which, despite spectral filtering, influence observation of single NV centers as background.   \\
We study the SiV ensemble in detail focusing on the fine structure of the SiV zero phonon line (ZPL) which is a meaningful measure of the crystalline quality of the diamond as strain in the material shifts and broadens the ZPL fine structure components in an ensemble. Consequently, we measure the ZPL fine structure at $10\,$K in different depths along the growth direction (see Fig.\ \ref{fig:SiV}). We observe a clear fine structure close to the growth surface while ZPL broadening completely masks the fine structure deeper in the diamond towards the nucleation side where the growth started. We expect the ZPL to broaden linearly with dislocation density.\cite{Hizhnyakov1999} To verify this fact, we fit the linewidth using a 1/layer thickness$^n$ dependence. We obtain $n\approx 1$ which is in  good agreement with the predicted dislocation density evolution during growth which should lead to an estimated density of dislocations of $\approx$ $10^7\,$cm$^{-2}$ close to the growth surface.\cite{Schreck2017} In this area, we observe a fitted linewidth of $61.2(5)\,$GHz (details see supplementary material). Our observed linewidths are thus only roughly a factor of 6 higher than in high-quality, thin, homoepitaxial diamond films \cite{Neu2013} witnessing high crystalline quality.
\\
To test the applicability of the material for NV-based quantum technologies, we investigate single, native NV centers in the highest-quality region namely the first micrometers below the growth surface. To prove the observation of single NV centers, we measure second order photon correlations $g^{(2)}$ [see Fig.\ \ref{fig:NVsingle}(c)]. We achieve a signal-to-background ratio of less than 4, limited by SiV PL. Nevertheless, these measurements clearly show single photon emission of NV centers, while the SiV ensemble PL limits $g^{(2)}(0)$ (see supplementary material). From several PL maps [see Fig. \ref{fig:NVsingle}(a)], we determine an NV density of $\approx$ 1 NV per $\mu$m$^3$ equivalent to a concentration of $\approx 0.005\,$ppb. This value is in reasonable accordance with the density estimated from the nitrogen impurity concentration in the feed gases (details see supplementary material). Subsequently, we perform photoluminescence lifetime imaging in areas with clearly discernible, individual NV centers (data see supplementary material). The NV centers show on average a lifetime of $12.3(6)\,$ns which agrees well with the lifetime in bulk diamond.\cite{Manson2006} This finding rules out quenching e.g.\ due to structural defects associated with the incorporation of silicon impurities\cite{Bolshakov2015} as well as quenching via potential near-field energy transfer between NV and SiV centers.\cite{Monticone2013} Furthermore, we perform ground state ODMR measurements of single NVs. We obtain the ODMR-spectrum in Fig.\ \ref{fig:NVsingle}(d) with narrow resonances (external magnetic field of $11\,$G applied) and a contrast exceeding $10\,\%$. The zero-field-splitting of $2.876\,$GHz indicates low strain close to the growth surface which would otherwise shift or split the resonance. Spin-Echo measurements of single NV centers show a coherence time of $T_2 = 5(1)\,\mu$s.
\\
Efficient nanoscale sensing using color centers in diamond typically requires shallow NV centers ($< 50\,$nm below the surface) incorporated into dedicated nanostructures e.g.\ nanopillars.\cite{Appel2016} To test the usability of our novel diamond material, we shallowly implant nitrogen ions with an implantation energy of $6\,$keV and doses of $1.5$ and $3\times10^{11}\,$cm$^{-2}$ to form an NV layer $10\,$nm below the diamond surface. After annealing the sample, PL maps reveal homogenous creation of the NV layer across the sample. PL spectra from the implanted layer [see Fig.\ \ref{fig:NVensemble}(a)] show strong NV PL with a minor part of PL originating from neutral NV centers; strong SiV PL is still visible. Due to the enhanced number of NVs, their PL is more pronounced, whereas the SiV PL remains mainly unchanged. The PL lifetime of the implanted NV centers is longer than the native NV centers' lifetime due to their proximity to the surface. We find $\tau = 17\,$ns, typical for shallowly-implanted NV centers.\cite{Appel2016} To demonstrate the usability of the NV center ensemble for magnetic sensing, we measure ODMR without applying an external magnetic field which shows a contrast of $C=20\,\%$ [see Fig.\ \ref{fig:NVensemble}(b)] at $2.871(2)\,$GHz. Power broadening for the ODMR resonance occurs because of the strong microwave field necessary to saturate the ground state spin transition in the ensemble. To estimate the unbroadened linewidth, we follow Ref.\ \onlinecite{Dreau2011} and obtain a linewidth of $10.67\,$MHz (see supplementary material). Rabi oscillations of a sub-ensemble of the implanted NV centers using a bias magnetic field of $38\,$G confirm coherent manipulation of the spins. The spin coherence time $T_2 = 5.2(3)\,\mu$s of the implanted NV centers is comparable to $T_2$ for the native NVs [see Fig.\ \ref{fig:NVensemble}(c)]. Consequently, we assume that $T_2$ is limited by the properties of our diamond material. Often, the concentration of substitutional nitrogen [N]$^s$ limits  $T_2$.\cite{Rondin2014} However, as the estimated concentration of [N]$^s$ is in the ppb range for our diamond material, we would expect $T_2$ to be at least an order of magnitude higher than observed. We also exclude minor concentrations of compensated boron acceptors as the source of decoherence. A possible spin bath reducing $T_2$ for the observed native NV centers might arise from various paramagnetic, silicon-containing defects in the material including neutral SiV centers and optically-inactive silicon-hydrogen complexes.\cite{DHaenens2010} Furthermore, vacancy complexes especially di-vacancies have been identified as a source of decoherence for implanted NV centers recently.\cite{deOliveira2017} We tentatively suggest that vacancy complexes forming at dislocations might reduce $T_2$ in our material. 
\\
Another important test for the material is to manufacture well-defined and stable nanostructures by reactive ion etching like, e.g., nanopillars or scanning probes.\cite{Appel2016} Such photonic nanostructures significantly enhance photon rates from single color centers, consequently boosting sensor sensitivity.\cite{Fuchs2018} We use the pristine growth surface ($2.2\,$nm rms roughness, area $2\,\mu$m$\times$$2\,\mu$m) to etch nanopillars in an inductively coupled reactive ion etching plasma (Ar/O$_2$, $50\,$sccm each, $18.9\,$mTorr, $500\,$W ICP, $200\,$W RF power) enabling highly-anisotropic etching of diamond. The nanopillars have an almost cylindrical shape with a diameter of $\approx 200\,$nm and a height of $\approx 1\,\mu$m [see Fig.\ \ref{fig:NVensemble}(d)] closely resembling scanning probes optimized for high sensitivity magnetometry.\cite{Fuchs2018} The surface roughness remains mainly unchanged ($<4\,$nm rms roughness, area $2\,\mu$m$\times$$2\,\mu$m) with no etchpits forming. We note that considering the estimated density of dislocation, the probability to find a dislocation within an individual pillar is only $\approx 1\%$ indicating the suitability of our high quality heteroepitaxial diamond material for the production of quantum technology devices. Thus, our wafer-sized diamonds open up the way for up-scaling nanofabrication of NV center-related structures for various applications.
\\
In conclusion, we show that our wafer-sized, heteroepitaxial single-crystal diamonds contain single, native NV centers with a density of roughly 1 NV per $\mu$m$^3$ embedded in an ensemble of SiV centers. Native NVs show moderate coherence time of $T_2 = 5\,\mu$s and are suitable for ODMR-related sensing applications. Low-temperature spectroscopy reveals a clear fine structure of the SiV ZPL close to the growth surface witnessing low strain and high crystalline quality. Additionally, we shallowly implant NV centers as required for nanoscale sensing  and demonstrate the fabrication of nanopillars into the pristine growth surface of the diamond. \\
The outstanding advantage of the present material system is its commercial availability and superior size, potentially enabling up-scaling of nanostructure fabrication for sensing and consequently reducing manufacturing cost for diamond-based devices. Due to wafer-scale growth, also the cost per carat can be reduced by potentially one order of magnitude compared to present high-purity, single crystal diamond. Consequently, the material has the potential to eliminate a bottle neck in diamond-based technologies. Challenges arise from moderate NV coherence times in the material and the detrimental influence of the SiV PL on NV spin readout limiting the usability of NV centers for sensing. To reduce the incorporation of silicon, we envisage overgrowing millimeter-thick diamond wafers after removal of the Ir/YSZ/Si substrate since the Si-wafer is the main source of silicon in the process. For many applications, only thin active layers ($<10\,\mu$m) are required.  As a consequence, a slow growth rate and process conditions for optimized crystal quality can be chosen. We anticipate that this approach will considerably improve NV coherence in the active zone.  Nevertheless, the observed native NV centers are already useful for sensing using ODMR resonances shifts to detect magnetic fields with moderate strength e.g.\ when performing failure analysis of electric circuits \cite{Nowodzinski2015} or imaging of domain walls in thin ferromagnetic films applicable in spintronics.\cite{Tetienne2014} Similarly, implanted ensembles of NV centers are suitable for, e.g., wide-field imaging of magnetic particles in living cells.\cite{LeSage2013}
\\

For supplementary information about nitrogen incorporation, background determination of the $g^{(2)}$-measurements, fitting of the SiV PL, PL lifetime maps and estimation of the power broadening, see [APL URL].\

This research has been funded via the NanoMatFutur program of the German Ministry of Education and Research (BMBF) under grant number FKZ13N13547. We acknowledge funding via a PostDoc fellowship of the Daimler and Benz foundation. We thank J\"org Schmauch for recording SEM images as well as Dr.\,Rene Hensel for use of the ICP-RIE and assistance.

\bibliography{Paper}

\end{document}